\documentclass{article}

\usepackage{amssymb,amsmath,bm,graphicx,multirow}
\def\rd{\mathrm{d}}

\begin{document}
\begin{titlepage}
\begin{center}
\noindent{\large\textbf{Making sessile drops easier}}
\vspace{2\baselineskip}

Amir H. Fatollahi$^{~(1}$
~~~and~~~Maryam Hajirahimi$^{~(2}$
\vspace{0.7cm}

\textit{1)  Department of Physics, Alzahra University, \\ P. O. Box  19938, Tehran  91167, Iran\\
email: fath@alzahra.ac.ir
 \\
\vspace{0.5cm}
2) Physics Group, South Tehran Branch, Islamic Azad University, \\ P. O. Box 11365, Tehran 4435, Iran,\\
email: m\_hajirahimi@azad.ac.ir
}
\end{center}
\vspace{\baselineskip}
\begin{abstract}
Using an identity, directly derived from the Young-Laplace equation,
the problem of the equilibrium shape of an axisymmetric sessile drop is reduced to a
one-parameter shooting method problem. Based on the method the numerical
solutions for drops with Bond number up to 15 are plotted.
The agreement between the method and the ADSA-D method
as well as the experimental data is tested. A \texttt{Mathematica} code
based on the method is presented.
\end{abstract}

\vspace{1cm}


\end{titlepage}

\section{Introduction}
The problem of a drop on a horizontal surface with the effect of surface tension being balanced with gravity has been studied for more than a century. The early numerical solutions go back to 1883 \cite{adams}, with updates by different authors \cite{staicop,padday,hartland}.
Different perturbative treatments of the problem have been developed over the years,
among them are those by \cite{chester,ehrlich,smith,shan82,shan84} for small drops (small Bond number). As large drops (or vanishing surface tension drops) are theoretically an infinitely large and thin film of liquid subjected to the boundary conditions at the outer edge, the limit of large Bond number falls and has been studied in the context of singular perturbation problems \cite{rienstra}. Based on the similarity between the truncated oblate spheroid and drop's shape, an approximated profile is suggested in \cite{ryley} for the shape of the drop.
In \cite{lehman} a new numerical treatment of the problem is given based on a variational method to minimize the total energy of the drop, by which the use of the tables by \cite{adams} is more direct than the earlier treatments. As another effort in this direction, in \cite{obrien} the singular perturbation technique is used to obtain the asymptotic expressions describing the shape of small sessile and pendant drops.
The study of the profiles of resting drops in different situations is particularly important for practical purposes. In fact, one of the most common methods to measure the surface tension of liquids is based on the matching between calculated drop's profiles and measured drop's shapes.
Over the years, the optimization of matching methods between the calculated profiles and the experimental data on drop's profiles has been the subject of several research pieces \cite{maze, neumann,kwok}.

Treating a sessile drop as a boundary value problem, it is shown that the multi-parameter shooting
method is applicable (see \textit{e.g.} \cite{graham,neum97,kwok91}). It is the purpose of this note to show that a one-parameter shooting method is applicable to sessile drops as well. The reduction of parameters is the result of using an identity directly derived from the
Young-Laplace equation \cite{fathscr}. By this identity the numerical procedure to reach the solution can be controlled by only one shooting parameter.

The scheme of the rest of this note is as follows. In Sec.~2 the basic notions are shortly reviewed.
In Sec.~3 the mathematical setup as well as the derivation of the mentioned identity is presented. Sec.~4 is devoted to some results and comparisons with some available numerical and experimental data. In Appendix a code in \texttt{Mathematica} based on the developed shooting procedure is presented.

\section{Basic notions}
The shape of a drop of liquid on a solid surface, in the idealized case (absence of impurities and pinning effects), is determined by the quantities:
1) the surface tension of liquid $\gamma$, 2) the solid-liquid adhesion
coefficient $\sigma$, 3) the shape of the solid surface, and due to the weight, 4) the drop's volume.
At the solid-liquid-vapor (s.l.v.) point of contact, the contact angle $\vartheta$ in the equilibrium condition is given
by the Young equation
\begin{equation}\label{1}
\cos \vartheta =  \frac{\sigma}{\gamma} -1.
\end{equation}
Three classes of possibilities for the contact angle are presented in Fig.~1.

\begin{figure}[t]
\begin{center}
\includegraphics[width=1.0\columnwidth]{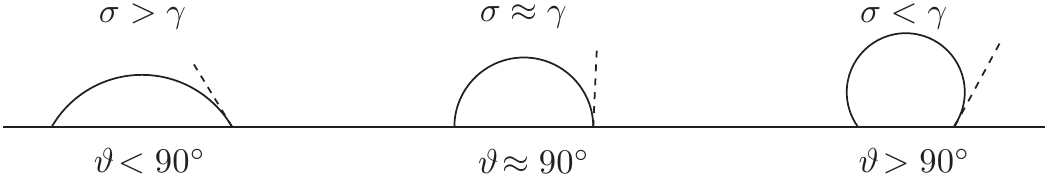}
\caption{Three classes of the drop's shape on a solid surface. }
\end{center}
\end{figure}

At every point on the drop surface the Young-Laplace relation holds
\begin{equation}\label{2}
  \gamma \,\bigg (\frac{1}{R_1} + \frac{1}{R_2} \bigg)=\Delta p
\end{equation}
\noindent in which $\Delta p\equiv p_\mathrm{l}-p_\mathrm{v}$
is the pressure jump across the surface, and $(R_1,R_2)$ are two principal
radii of curvature of the surface at the point. Provided by the hydrostatic laws, $\Delta p$ is expressed in terms of the surface equation. Hence, the Young-Laplace relation is the partial differential equation which, accompanied by appropriate boundary conditions, determines the shape of the drop's surface.

Heuristically, the final shape of drop is the result of the balance between the surface effects and the bulk ones. While the surface tension tends to decrease the surface of the drop, the adhesion coefficient tends to increase the surface of the contact region, and the gravity tends to lower the center of mass of the drop. In many practical cases the surface tension and the adhesion coefficient, though with opposite effects, may be considered at the same order, meaning $\gamma$ and $\sigma$ are comparable.
For a drop with volume $V$ and density $\varrho$, the so-called Bond number defined by the
dimensionless combination $V^{2/3}\varrho g/\gamma$ would determine whether weight has the dominant contribution or not. When the weight is ignorable, only the contribution from the surface effects exist. Minimizing the area for a fixed volume, the drop's surface is part of a sphere.

\section{The mathematical setup}
Using the cylindrical coordinate setup given in Fig.~2, the total curvature of a surface
with azimuthal symmetry, represented by $z=f (\rho) $ is given by \cite{fathscr}
\begin{equation}\label{3}
\frac{1}{R_1} + \frac{1}{R_2} = \frac{1}{\rho}~\frac{\rd }{\rd \rho}\bigg(\rho \,
\frac {|f '|} {\sqrt {1 + f '^ {\, 2}}} \bigg),
\end{equation}
\noindent where $ f '=\rd  f/\rd  \rho $. The pressure jump in presence of gravity gets contribution from the weight of the drop's layers as well, leading to
\begin{equation}\label{4}
\Delta p (z) =\Delta p_\gamma +  \varrho g (h - z)
\end{equation}
in which $h$ is the height of the drop's apex, and $\Delta p_\gamma$ is a constant
representing the pressure jump due to the
surface tension. So, the Young-Laplace  relation reads
\begin{equation}\label{5}
\mp\frac{1}{\rho} \frac{\rd }{\rd  \rho} \bigg(\rho \frac {f '_\pm} {\sqrt {1 + f '^ {\, 2}_\pm}} \bigg) =  2 \kappa +  \frac{\varrho g}{\gamma} (h- f_\pm).
\end{equation}
in which $f_+$ and $f_-$ are denoting the upper and lower parts of the drop, respectively (see Fig.~2b), and $\kappa:=\Delta p_\gamma/(2\gamma)$. At apex ($h=f_+(0)$) we have $R_1=R_2$, and so by (\ref{2}), $\kappa$ is simply the curvature at apex of drop. We mention $f'_+<0$ and $f'_->0$. The boundary conditions for $\vartheta>90^\circ$ are:
\begin{align} \label{6}
f'_+(0)&=0\\ \label{7}
f'_-(\rho_0)&=-\tan\vartheta,\\ \label{8}
f_-(\rho_0)&=0.
\end{align}
In case with $\vartheta<90^\circ$ (\ref{7}) and (\ref{8}) are valid for $f_+$.

\begin{figure}[t]
\begin{center}
\includegraphics[width=1.0\columnwidth]{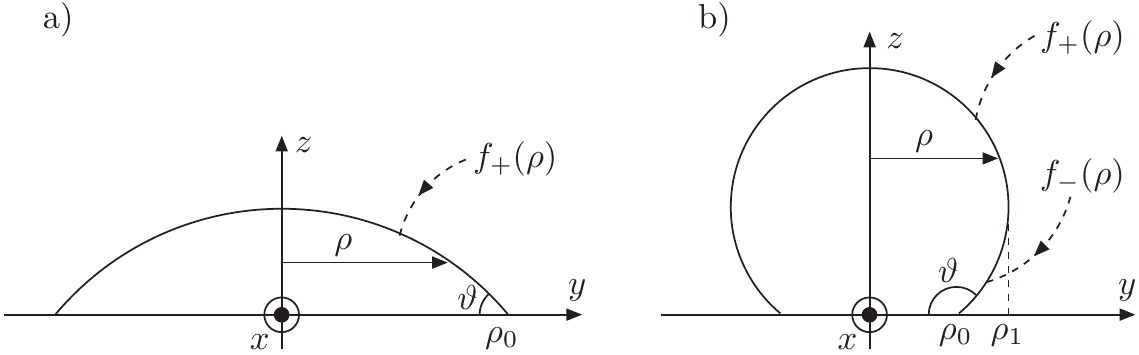}
\caption{The geometry of the mathematical setup for: a) $\vartheta < 90^\circ$, b)
$\vartheta > 90^\circ$. }
\end{center}
\end{figure}

The main issue with equation (\ref{5}) is that the parameters $h$ and $\kappa$ are not known at the first place, and would be determined only after the complete solution is available. So at starting point the main equation is not fully known. Further, the contact radius $\rho_0$
(Fig.~2), as the limiting value for the variable $\rho$, is not known at first place.
These all indicate that the Young-Laplace equation can not be treated as straightforward as
an ordinary boundary value problem.
As will be seen shortly, by integrating the Young-Laplace equation an identity is obtained
which relates the three unknown parameters in a very helpful way.
In what follows we mainly consider the case with $\vartheta > 90^\circ$. The generalization to case with $\vartheta < 90^\circ$  is rather straightforward.
Integrating the Young-Laplace equation for the upper and lower parts of drop leads to \cite{fathscr}
\begin{align}\label {9}
 \rho_1 &=    \left( \kappa +\frac{\varrho g}{2\gamma}h\right) \rho_1 ^ 2 -
\frac {\varrho g} {\gamma} \int_ {0} ^ {\rho_1} \rho f_+(\rho) \rd  \rho \\
\label{10}
\rho_1- \rho_0 \sin \vartheta  & =  \left( \kappa +\frac{\varrho g}{2\gamma}h\right)
(\rho_1 ^ 2 - \rho_0 ^ 2) - \frac {\varrho g} {\gamma} \int_ {\rho_0} ^ {\rho_1} \rho f_-(\rho)\rd  \rho
\end{align}
in which we have used
\begin{equation}\label{11}
\left. \frac  {f _-'} {\sqrt {1 + f_-'^{2}}} \right|_{\rho_0}\!\!\!\! =
\frac {-\tan \vartheta } {\sqrt {1 + \tan ^ 2 \vartheta }} = \sin \vartheta
\end{equation}
for $\vartheta > 90^\circ$. Subtracting (\ref{9}) and (\ref{10}) gives \cite{fathscr}
\begin{equation}\label{12}
\kappa +\frac{\varrho g}{2\gamma}h
= \frac{\sin \vartheta }{\rho_0} + \frac {\varrho g V} {2 \pi \gamma \,\rho_0^2}
\end{equation}
in which we have used the relation for the volume of drop,
\begin{equation} \label {13}
\frac  {V} {2 \pi}  = \int_0 ^ {\rho_1} \rho f_ +(\rho) \rd  \rho - \int_ {\rho_0} ^ {\rho_1} \rho f_- (\rho)\rd  \rho.
\end{equation}
\noindent  It is easy to show that identity (\ref{12}) is valid for the acute contact angle
($\vartheta < 90^\circ$) as well. It is reminded that in obtaining (\ref{12}) no approximation is used, and so it is an exact relation.

Now, using the identity (\ref{12}), the combination $\kappa$ and $h$ in right-hand side of (\ref{5}) can be replaced in favor of $\rho_0$, and the Young-Laplace equation turns to:
\begin{equation}\label{14}
\mp\frac{1}{\rho} \frac{\rd }{\rd  \rho} \bigg(\rho \frac {f '_\pm} {\sqrt {1 + f '^ {\, 2}_\pm}} \bigg) =
2\left(\frac{\sin \vartheta }{\rho_0} + \frac {\varrho g V} {2 \pi \gamma \,\rho_0^2}\right)
-\frac{\varrho g}{\gamma}  f_\pm.
\end{equation}
In above the only unknown parameter is $\rho_0$. For the case with $\vartheta < 90^\circ$ only $f_+$ in above should be kept.

For later use, let us remind the spherical solution of weightless drop. It is easy to check that, by setting $g=0$ and $\kappa=1/R$, the expression \cite{fathscr}
\begin{equation}\label{15}
z = f_{0\pm} (\rho) =z_0 \pm \sqrt {R^2 - \rho ^ 2} ,
\end{equation}
\noindent satisfies (\ref{5}). The radius $R$ is fixed by the volume of spherical cap,
\begin{equation}\label{16}
V=\frac{\pi}{3} R^3\, (1-\cos\vartheta)^2(2+\cos\vartheta).
\end{equation}
Following a simple geometrical argument in the sphere (Fig.~2), we have
\begin{align}\label{17}
\left. \begin{matrix}
& \rho_0=R\sin\vartheta \cr
& z_0=-R\cos\vartheta \cr
& \rho_1=R \cr
& h=R+z_0
\end{matrix}\right\} ~~~~ g=0
\end{align}

\begin{figure}[t]
\begin{center}
\includegraphics[width=0.6\columnwidth]{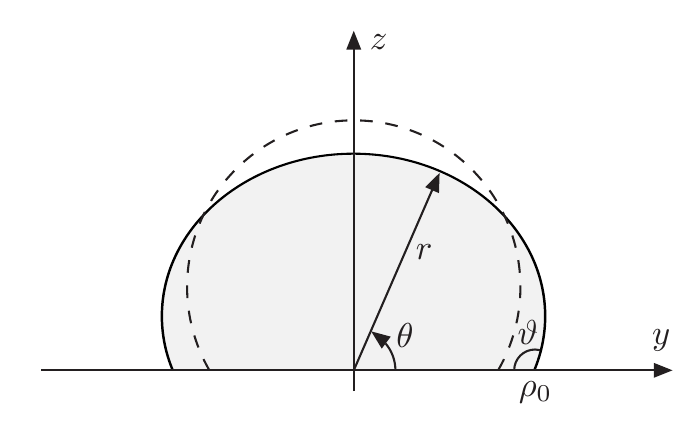}
\caption{The geometry with the polar coordinate $(r,\theta)$. The dashed curve is the sphere solution of weightless drop. }
\label{fig3}
\end{center}
\end{figure}

As in case with $\vartheta > 90^\circ$ around the equatorial radius $\rho_1$ (Fig.~2b) $f'_\pm\to\mp\infty$, it would be convenient to switch to the spherical coordinates in which the whole surface of drop is covered by one function. Among many others, we found the choice illustrated in Fig.~3 more convenient, in which
\begin{align}\label{18}
z&=r(\theta)\,\sin\theta,\\  \label{19}
\rho&= r(\theta)\,\cos\theta.
\end{align}
In the new coordinates the Young-Laplace equation takes the following form
\begin{align}\label{20}
\frac{\gamma}{r^2\cos\theta} \left[ \frac{\rd}{\rd \theta} \left( \frac{r\,r'\cos\theta}{\sqrt{r^2+r'^2}}\right)
-\frac{2 r^2+r'^2}{\sqrt{r^2+r'^2}}\cos\theta\right]=
\Delta p_\gamma+ \varrho g (h- z),
\end{align}
in which $r'=\rd r/\rd\theta$. Using the identity (\ref{12}) leads to
\begin{align}\label{21}
\frac{\rd}{\rd \theta} \left( \frac{r\,r'\cos\theta}{\sqrt{r^2+r'^2}}\right)
-\frac{2 r^2+r'^2}{\sqrt{r^2+r'^2}}\cos\theta=
\left[c\, r\sin\theta-
2\left(\frac{\sin \vartheta }{\rho_0} +
\frac {c\, V} {2 \pi \,\rho_0^2}\right)\right]r^2\cos\theta ,
\end{align}
in which $c=g\varrho/\gamma$, the so-called the capillary constant.
As we are considering axisymmetric drops, it is sufficient to restrict the
polar coordinate $\theta$ only to the left-side of drop,
\begin{align}\label{22}
0\leq \theta \leq 90^\circ.
\end{align}
Using (\ref{6})-(\ref{8}) and (\ref{18})-(\ref{19}) the
boundary conditions in new coordinates are found
\begin{align}
\label{23}
r'(0)&=-\rho_0 \cot\vartheta,\\
\label{24}
r(0)&=\rho_0,\\
\label{25}
r'(90^\circ)&=0.
\end{align}
Therefore the shooting method by one controlling parameter $\rho_0$ can be employed
to obtain numerical solutions of (\ref{21}). The value of $\rho_0^{g=0}=R\sin\vartheta$ of weightless drop can be chosen as the initial guess of the shooting parameter $\rho_0$. As it is known that, due to weight, the radius of contact region increases (see \textit{e.g.} \cite{fathscr}), upon stepwise increasing of the parameter $\rho_0$ and checking the condition (\ref{25}) at apex of drop, one can reach the desired accuracy.

\begin{figure}[h]
\begin{center}
\includegraphics[width=0.6\columnwidth]{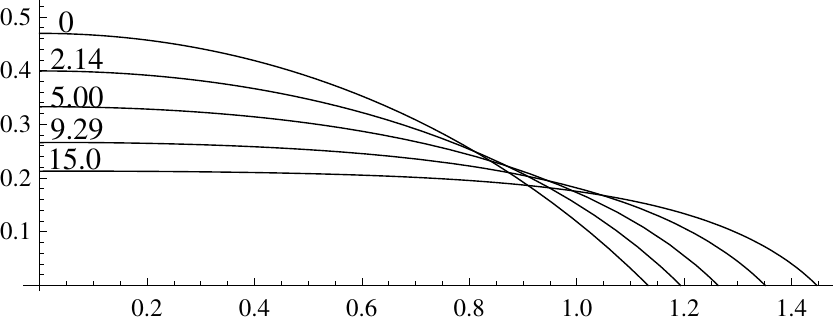}
\caption{The plots of solutions of (\ref{21}) for drops with acute contact angle but different Bond numbers (given on each plot). In all samples: $V=1$, $\gamma=70$, $\varrho=1$, and $\vartheta=45^\circ$ (the varying input value is $g$). Scales on axes: 1:1.}
\label{fig4}
%
\includegraphics[width=0.4\columnwidth]{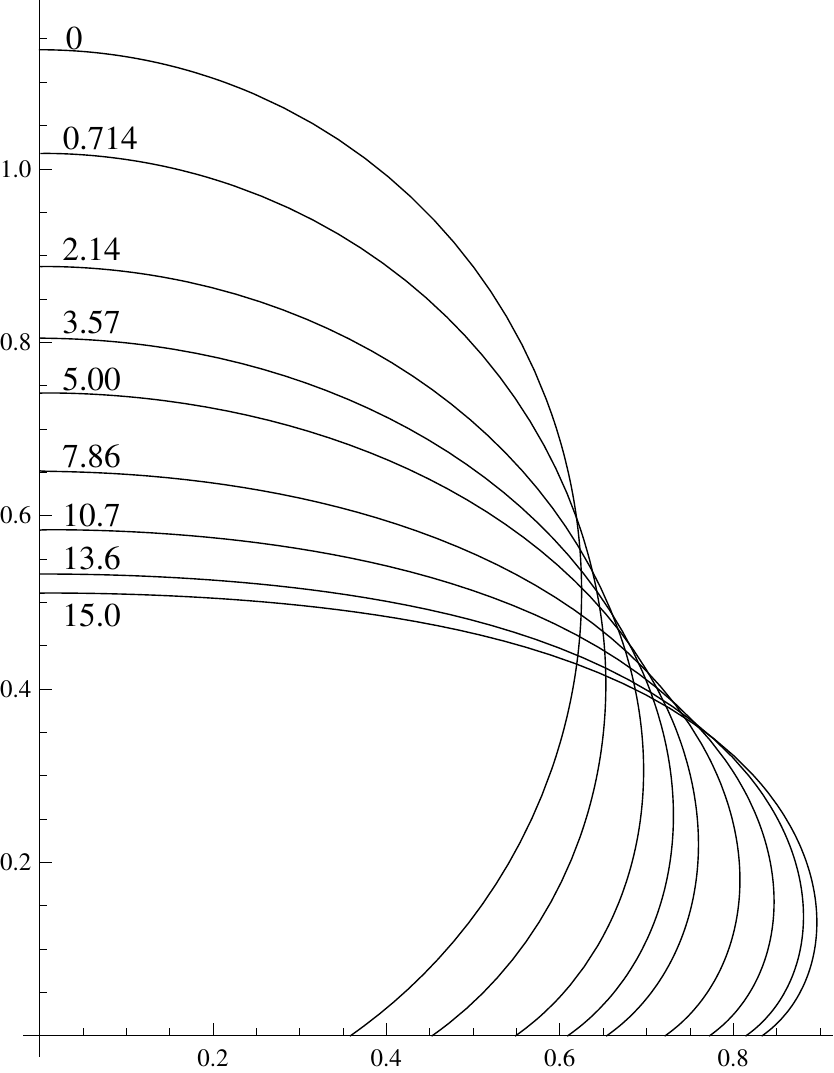}
\caption{The same as Fig.~4 but with obtuse contact angle $\vartheta=145^\circ$. }
\label{fig5}
\end{center}
\end{figure}

\section{Samples of solutions and comparisons}
In order to assess the accuracy of the present method, the results based on it are compared with some available data. However, it would be illustrative to begin with a demonstration of the results based on the method. The numerical solutions of (\ref{21}) with different Bond numbers for $\vartheta =45^\circ$ and $\vartheta =145^\circ$ are presented in Fig.~4 and Fig.~5, respectively.
The presented plots cover Bond numbers from 0 to 15. As expected, by increasing the Bond number, the apex's height decreases while contact radius, as well as equatorial radius for case with $\vartheta>90^\circ$, increase.

As a further test, in Tab.~1 the results of application of one-parameter shooting method to (\ref{21}) are compared with those by \cite{neum97} by the ADSA-D method. The input values used by \cite{neum97} are taken from the data generated by ALFI program. It is mentioned that the present method returns the correct values with negligible errors.

\begin{table}[h]{\scriptsize
\begin{center}
\begin{tabular}{c c c  c c | c c }
 \multicolumn{5}{c}{Reported}  & \multicolumn{2}{c}{This work}  \\
Drop/$\vartheta$ & $c$   &  $V$ &  $\kappa$  & $R_\mathrm{max.}$ & $\kappa$  & $R_\mathrm{max.}$ \\
 & cm$^{-2}$ & cm$^{3}$ & cm$^{-1}$ & cm & (err.) & (err.) \\
\hline
& & & & & & \\
S1/$120^\circ$ & 13.45 & 0.4578 & 0.7340 & 0.6562 &  0.7338  & 0.6562 \\
& & & & & (0.03\%) & (0.00\%) \\
S2/$10^\circ$ & 13.45 & 0.1236 & 0.030 & 1.1152 &  0.030  & 1.1150 \\
& & & & & (0.00\%) & (0.02\%) \\
S3/$80^\circ$ & 13.45 & 0.0014 & 10.0 & 0.0965 &  10.09  & 0.0957 \\
& & & & & (0.90\%) & (0.84\%) \\
S4/$30^\circ$ & 1.00 & 0.0486 & 1.00 & 0.4849 &  1.00  & 0.4846 \\
& & & & & (0.00\%) & (0.06\%) \\
S5/$45^\circ$ & 1000 & 0.00089 & 1.00 & 0.1299 &  1.00  & 0.1299 \\
& & & & & (0.00\%) & (0.00\%)
\end{tabular}
\caption{{\small The comparison between the numerical solutions by \cite{neum97} and those by the present work. The values by \cite{neum97} are taken from Tabs.~1 \& 2. The parameter $c$, as in this work, denotes the capillary constant $g\varrho/\gamma$. The curvature at apex ($\kappa$) is denoted in \cite{neum97} by $b$.  $R_\mathrm{max.}$ denotes the contact radius $\rho_0$ for $\vartheta<90^\circ$ and the equatorial radius $\rho_1$ for $\vartheta>90^\circ$ (Fig.~2).}}
\end{center}
}\end{table}

The comparison with some available experimental data with the values by solutions of (\ref{21}) are presented in Tab.~2 and Tab.~3 for acute and obtuse contact angles, respectively.

\begin{table}[h]{\scriptsize
\begin{center}
\begin{tabular}{c c c  c c | c c }
 \multicolumn{5}{c}{Reported}  & \multicolumn{2}{c}{This work}  \\
Drop & $\vartheta$   &  $V$ &  $\rho_0$  & $h$ & $\rho_0$  & $h$ \\
Specification &  & $10^{-3}$cm$^{3}$ & cm & cm & (err.) & (err.) \\
\hline
& & & & & & \\
Water on & $72^\circ$ & 6.75 & 0.1748 & 0.1148 & 0.1741 &  0.1199   \\
carbon& & & & & (0.4\%) & (4.4\%) \\
steel \cite{ryley}  & $71.3^\circ$ & 13.5 & 0.2240 & 0.1411 & 0.2225 &  0.1469   \\
& & & & & (0.7\%) & (4.1\%) \\
\hline
& & & & & & \\
Water on & $73.44^\circ$ & 123.4 & 0.4897 & -  & 0.4891 &  0.2662   \\
PMMA \cite{kwok}& & & & & (0.1\%) &  \\
\hline
& & & & & & \\
 & $76^\circ$ & $10$ & 0.388/2 & 0.141  & 0.195 & 0.138    \\
Formamide& & & & & (0.5\%) & (2.1\%)  \\
on PE \cite{shan82} & $76.5^\circ$ & $40$ & 0.643/2 & 0.208  & 0.3203 & 0.2015    \\
& & & & & (0.4\%) & (3.1\%)  \\
& $76.5^\circ$ & $100$ & 0.890/2 & 0.253  & 0.454 & 0.245    \\
& & & & & (2.0\%) & (3.2\%)
\end{tabular}
\caption{{\small The comparison between experimental data for drops with $\vartheta<90^\circ$ and values by numerical solutions of (\ref{21}).
For water drops: $\varrho=0.997~\mathrm{g/cm}^3$, $\gamma=72.0~\mathrm{mJ/m}^2$. For formamide drops:
$\varrho=1.133~\mathrm{g/cm}^3$, $\gamma=58.2~\mathrm{mJ/m}^2$.
For all drops: $g=980.7~\mathrm{cm/s}^2$.}}
\end{center}
}\end{table}

\begin{table}[h]{\scriptsize
\begin{center}
\begin{tabular}{c c c  c c c | c c c }
 \multicolumn{6}{c}{Reported}  & \multicolumn{3}{c}{This work}  \\
Drop & $\vartheta$   &  $V$ &  $\rho_0$ & $\rho_1$  & $h$ &  $\rho_0$  & $\rho_1$ & $h$ \\
Specification &  & $10^{-3}$cm$^{3}$ & cm & cm & cm & (err.) & (err.) & (err.) \\
\hline
& & & & & & \\
Water on & $117.34^\circ$ & 89.2 & - & 0.6728/2 & - & 0.3189 & 0.3371 & 0.3417   \\
coated mica& & & & & & & (0.2\%) &  \\
(FC-721) \cite{kwok91}& $117.63^\circ$ & 89.4 & - & 0.6735/2 & - & 0.3185 & 0.3371 & 0.3424   \\
& & & & & & & (0.1\%) &  \\
\hline
& & & & & & \\
& $136^\circ$ & 1.27 & 0.0493 & 0.0700 & 0.1103 & 0.0517 & 0.0699  & 0.1119 \\
Mercury & & & & & & (4.9\%) & (0.1\%)  & (1.5\%)  \\
on glass & $138^\circ$ & 3.16 & 0.0705 & 0.0965 & 0.1430 & 0.0717 & 0.0960  & 0.1476 \\
\cite{shan84} & & & & & & (1.7\%) & (0.5\%)  & (3.2\%)  \\
& $134^\circ$ & 9.00 & 0.1157 & 0.1435 & 0.1950 & 0.1160 & 0.1414  & 0.1930 \\
 & & & & & & (0.3\%) & (1.5\%)  & (1.0\%)  \\

\end{tabular}
\caption{{\small The comparison between experimental data for drops with $\vartheta>90^\circ$ and values by numerical solutions of (\ref{21}). For water drops: $\varrho=0.997~\mathrm{g/cm}^3$, $\gamma=72.0~\mathrm{mJ/m}^2$. For mercury drops: $\varrho=13.55~\mathrm{g/cm}^3$, $\gamma=480~\mathrm{mJ/m}^2$.
For all drops: $g=980.7~\mathrm{cm/s}^2$.
}}
\end{center}
}\end{table}

\appendix

\section{\texttt{Mathematica} code}
Here a code in \texttt{Mathematica} based on the method is presented.
In the below the following input values from Tab.~3 are taken:
$V=0.0892~\mathrm{cm}^3$, $\gamma=72.0~\mathrm{mJ/m}^2$,   $g=980.7~\mathrm{cm/s}^2$, $\varrho=0.997~\mathrm{g/cm}^3$,
$\vartheta=117.34^\circ$. The increasing step in $\rho_0$ is taken equal
to $0.00001$, with the criteria that slope at apex would be less than
$0.1$. The outputs of the code, together with the function
$r(\theta)$ (\texttt{ryl[tha]}), are: the resulting slope at the apex (\texttt{slope}),
the contact radius $\rho_0$ (\texttt{rh0}), the apex's height (\texttt{h}),
the angel $\theta_1$, radius $\rho_1$, and height $h_1$ at equator 
(\texttt{th1}, \texttt{rh1} and \texttt{h1}), together with the plot of the drop.

\vskip 1cm

{\fontfamily{cmtt}\selectfont
\noindent
{V=0.0892;{gamma}=72;g=980.7;{dens}=0.997;{vth}=117.34 {Degree};}\\
{R=(3 V/({Pi}(1-{Cos}[{vth}])${}^{\wedge}$2(2+{Cos}[{vth}])))${}^{\wedge}$(1/3);}\\
{c=g {dens}/{gamma};{rh0sph}=R {Sin}[{vth}];}\\
{{thmin}=0 {Degree};{thmax}=89.99 {Degree};}\\
{{ylsol}[{rh0$\_$}]{:=}{NDSolve}[\{D[r[{th}] r'[{th}] {Cos}[{th}]/{Sqrt}[r[{th}]${}^{\wedge}$2+r'[{th}]${}^{\wedge}$2],{th}]}\\
{-(2 r[{th}]${}^{\wedge}$2+r'[{th}]${}^{\wedge}$2){Cos}[{th}]/{Sqrt}[r[{th}]${}^{\wedge}$2+r'[{th}]${}^{\wedge}$2]}\\
{+2({Sin}[{vth}]/{rh0}+c V/(2 {Pi} {rh0}${}^{\wedge}$2))r[{th}]${}^{\wedge}$2 {Cos}[{th}]}\\
{-c r[{th}]${}^{\wedge}$3{Sin}[{th}] {Cos}[{th}]==0,}\\
{r[{thmin}]=={rh0},r'[{thmin}]==-{rh0} {Cot}[{vth}]\},r,\{{th},{thmin},{thmax}\}];}\\
{{slope}[{rh0$\_$}]{:=}r'[{thmax}]{/.}{ylsol}[{rh0}][[1]]{//}{Quiet};}\\
{{rh0t}={rh0sph};{While}[{Abs}[{slope}[{rh0t}]]>0.1,{rh0t}={rh0t}+0.00001];}\\
{{ryl}[{tha$\_$}]{:=}r[{tha}]{/.}{ylsol}[{rh0t}][[1]];}\\
{{Print}[{$\texttt{"}$slope=$\texttt{"}$},{slope}[{rh0t}]{//}N]}\\
{{apx}={ryl}[{thmax}]{//}N;}\\
{{Print}[{$\texttt{"}$rh0=$\texttt{"}$},{rh0t},{$\texttt{"}$ h=$\texttt{"}$},{apx}]}\\
{{th1}={tha}{/.}{FindRoot}[D[{ryl}[{tha}]{Cos}[{tha}],{tha}],\{{tha},0.05\}][[1]];}\\
{{rh1}={ryl}[{th1}]{Cos}[{th1}]{//}N;}\\
{{h1}={ryl}[{th1}]{Sin}[{th1}]{//}N;}\\
{{Print}[{$\texttt{"}$th1=$\texttt{"}$},{th1},{$\texttt{"}$ rh1=$\texttt{"}$},%
{rh1},{$\texttt{"}$ h1=$\texttt{"}$},{h1}]}\\
{{ParametricPlot}[\{{ryl}[{tha}] {Cos}[{tha}],{ryl}[{tha}]{Sin}[{tha}]\},\{{tha},{thmin},{thmax}\},}\\
{{AspectRatio}$\to${Automatic},{PlotStyle}$\to${Black}]}
}

\vskip 0.5cm
\textbf{Acknowledgement}:
The work by A.~H.~F. is supported by the Research Council of the Alzahra University.


\begin{thebibliography}{99}
\bibitem{adams} F. Bashforth and J.C. Adams, ``\textit{An attempt to test the theories of capillary attraction}", Cambridge University Press, Cambridge (1883).

\bibitem{staicop} D. N. Staicopolus, ``\textit{The computation of surface tension and of contact angle by the sessile-drop method}" (I and II), J. Colloid Interface Sci. \textbf{17} (1962) 439, \textit{ibid.} \textbf{18} (1963) 793.

\bibitem{padday} J. F. Padday, ``\textit{The profiles of axially symmetric menisci}", Phil. Trans. R. Soc. Lond. A \textbf{269} (1971) 265.

\bibitem{hartland} S. Hartland \& R. W. Hartley, ``\textit{Axisymmetric fluid-liquid interfaces}", Elsevier, Amsterdam (1976).

\bibitem{chester} A. K. Chesters, ``\textit{An analytical solution for the profile and volume of a small drop or bubble symmetrical about the vertical axis}", J. of Fluid Mech. \textbf{81} (1977) 609.

\bibitem{ehrlich} R. Ehrlich, ``\textit{An alternative method for computing contact angle from the dimensions of a small sessile drop}", J. Colloid Interface Sci. \textbf{28} (1968) 5.

\bibitem{smith} P. G. Smith \& T. G. M. van de Ven, ``\textit{Profiles of slightly deformed axisymmetric drops}", J. Colloid Interface Sci. \textbf{97} (1984) 1.

\bibitem{shan82} M. E. R. Shanahan, ``\textit{An approximate theory describing the profile of a sessile drop}", J. Chem. Soc., Faraday Trans. I \textbf{78} (1982) 2701.

\bibitem{shan84} M. E. R. Shanahan, ``\textit{Profile and contact angle of small sessile drops}", J. Chem. Soc., Faraday Trans. I \textbf{80} (1984) 37.

\bibitem{rienstra} S. W. Rienstra, ``\textit{The shape of a sessile drop for small and large surface tension}", J. Eng. Math. \textbf{24} (1990) 193.

\bibitem{ryley} D. J. Ryley and B. H. Khoshaim, ``\textit{A new method of determining the contact angle made by a sessile drop upon a horizontal surface (sessile drop contact angle)}", J. Colloid Interface Sci. \textbf{59} (1977) 243.

\bibitem{lehman} W. M. Robertson and G. W. Lehman, ``\textit{The shape of a
sessile drop}", J.  Appl. Phys. \textbf{39} (1968) 1994.

\bibitem{obrien} S. B. G. O'Brien, ``\textit{On the shape of small sessile
and pendant drops by singular perturbation techniques}", J. Fluid Mech. \textbf{233}
(1991) 519.

\bibitem{maze} C. Maze \& G. Burnet, ``\textit{A non-linear regression method for
calculating surface tension and contact angle
from the shape of a sessile drop}", Surface Sci. \textbf{13} (1969) 451.

\bibitem{neumann} Y. Rotenberg, L.~Boruvka \& A.~W.~Neumann, ``\textit{Determination of surface tension and contact angle from the shapes of axisymmetric fluid interfaces}", J. Colloid Interface Sci. \textbf{93} (1983) 169; P.~Cheng, D.~Li, L.~Boruvka, Y.~Rotenberg \& A.~W. Neumann, ``\textit{Automation of axisymmetric drop shape analysis for measurements of interfacial tensions and contact angles}", Colloids Surf. \textbf{43} (1990) 151.

\bibitem{kwok} D. Y. H. Kwok, ``\textit{Contact angles and surface energies}", Ph.D. Thesis, University of Toronto, page 32 (also available at http://www.mie.utoronto.ca/labs/last/kwok/drop.html).

\bibitem{graham} J. Graham-Eagle and S. Pennell, ``\textit{Contact angle calculations from the contact/maximum diameter of sessile drops}", Int. J. for Num. Meth. in Fluids \textbf{32} (2000) 851.

\bibitem{neum97} O. I. del Rio and A.~W.~Neumann, ``\textit{Axisymmetric Drop Shape Analysis: Computational Methods for the Measurement of Interfacial Properties from the Shape
and Dimensions of Pendant and Sessile Drops}", J. Colloid Interface Sci. \textbf{196} (1997)
136.

\bibitem{kwok91} E. Moy, P. Cheng, Z. Policova, S.~Treppo, D.~Kwok, D. R.~Mack,
P.~M.~Sherman, and A.~W.~Neumann, ``\textit{Measurement of contact angles from the maximum diameter of non-wetting drops by means of a modified axisymxnetric drop shape analysis}", Colloids Surf. \textbf{58} (1991) 215.

\bibitem{fathscr} A. H. Fatollahi, ``\textit{On the shape of a lightweight drop on a
horizontal plane}", Physica Scripta \textbf{85} (2012) 045401.

\end{thebibliography}
\end{document}